# Chapter 8

# Interface with Experiments

*H. Burkhardt and I. Efthymiopoulos*[*]

CERN, Accelerator & Technology Sector, Geneva, Switzerland

## 8      Interface with experiments

### 8.1    Introduction

The machine upgrade for high luminosity requires major changes on the machine side. Key ingredients for the luminosity increase are larger apertures in the focusing sections around the experiments and higher beam intensities. The experiments are upgraded for reduced inner beam pipes with more powerful vertex detectors. This is important for physics and essential for the increased pile-up.

Other key design considerations for the upgraded LHC detectors include longevity at increased radiation levels and minimization of activation. The definition of the machine-experiment interface issues and its timeline is described in the M18 document [1].

### 8.2    Interaction regions

Figure 8-1 shows the schematic layout of the LHC with its four interaction regions.

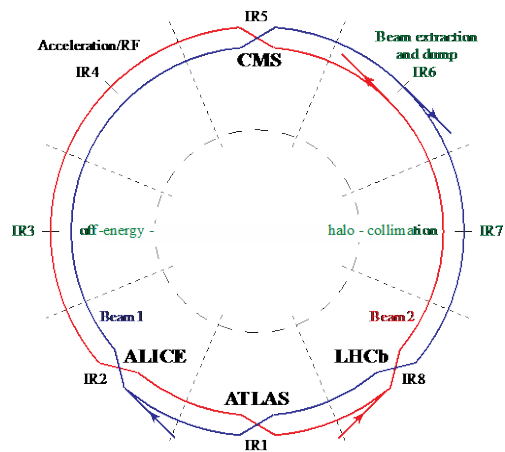

Figure 8-1: Schematic layout of the LHC with the four interaction regions that house the ALICE, ATLAS, CMS, and LHCb experiments.

The HL-LHC design is for four experiments: the two high luminosity experiments ATLAS and CMS at IR1 and IR5, respectively; ALICE at IR2; and LHCb at IR8.

Table 8-1 shows the target luminosities for the experiments in proton–proton operation for the original LHC design and for the high luminosity upgrade. Operation at so-called 'ultimate' parameters with peak luminosity of $7.5 \times 10^{34}$ cm$^{-2}$ s$^{-1}$ is considered. With levelling, this would lead to around 4000 fb$^{-1}$ by 2037.

---

[*] Corresponding author: ilias.efthymiopoulos@cern.ch

Table 8-1 Target luminosities for p–p operation for the HL-LHC. The luminosities for LHC Run 2 are also included for comparison. Total target integrated luminosity in CMS and ATLAS is 3000 fb$^{-1}$ after 12 years of operation.

| Experiment | IP | Peak levelled luminosity [cm$^{-2}$ s$^{-1}$] | |
|---|---|---|---|
| | | HL-LHC | LHC |
| ATLAS | 1 | $5 \times 10^{34}$ | $2 \times 10^{34}$ |
| CMS | 5 | $5 \times 10^{34}$ | $2 \times 10^{34}$ |
| ALICE | 2 | $1 \times 10^{31}$ | $1 \times 10^{31}$ |
| LHCb | 8 | $2 \times 10^{33}$ | $4 \times 10^{32}$ |

The main luminosity upgrade is for interaction regions IR1 and IR5, and will be implemented in the long shutdown LS3. The ALICE and LHCb experiments installed in IR2 and IR8 will have their most significant detector upgrades during LS2 scheduled for 2018/2019 and will continue to run after LS3. LHCb has requested a luminosity increase up to $2 \times 10^{33}$ cm$^{-2}$ s$^{-1}$. This is possible without changes to the magnet layout in IR8, and the required detector and vacuum beam pipe upgrades can be implemented in long shutdown LS2. It should be accompanied by improved shielding (including a minimal TAN upstream of D2), to minimize the impact of the increase in radiation and heating of cold machine elements. The low target luminosity for ALICE in p–p operation will require collisions with large transverse offsets.

The experimental programmes of other smaller experiments (LHCf, TOTEM) do not at present extend beyond LS3. Should these or other experimental proposals appear in the future, they will need to adapt to the HL-LHC beam conditions and installed hardware planned for the straight sections. The HL-LHC will be more constrained towards high luminosity/low $\beta$ operation than the present LHC. Very high $\beta^*$ running will have to end with LS3. The upper limit for $\beta^*$ for physics at top energy is expected to be reduced to about 30 m.

For HL-LHC operation, the machine layout in IR1 and IR5 will change significantly. The most relevant machine modification for the experiments will be the installation of the new large aperture triplet magnets Q1–Q3 in IR1 and IR5. Details are described in Chapter 3. What is important for the layout of the interface with the experiments is that the distance of the first quadrupole magnet (Q1) from the IP will remain the same (23 m) as before the upgrade. As a result, no modifications to the forward shielding of the experiments need to be implemented to accommodate changes in the accelerator layout. The inner coil diameter of the triplet magnets however will increase by roughly a factor of two from 70 mm to 150 mm. This implies that the inner beam pipe radius will have to increase between the triplet and the experiment. In particular, the passive absorbers (TAS) installed at 19 m from the interaction points will have to be replaced by a new larger aperture absorber (TAXS).

## 8.3 Experimental beam pipes

A key upgrade of the ATLAS and CMS experiments for HL-LHC operation is to the inner tracker detectors. The change to a smaller radius beam pipe was already made during LS1 [3, 4]. This is important for the detectors to deal with high pile-up and will be kept for HL-LHC operation. The ALICE and LHCb experiments also plan modifications to their experimental beam pipe to a reduced inner radius. The present and foreseen experimental beam apertures are summarized in Table 8-2.

The LHCb VELO is movable. It is only closed in stable physics to the value shown in the table, and retracted to 30 mm otherwise.

Table 8-2: Original and reduced inner beam pipe radii at the IPs

| IP | Experiment | When | Original $r_{min}$ [mm] | Reduced $r_{min}$ [mm] |
|---|---|---|---|---|
| 1 | ATLAS | LS1 | 29 | 23.5 |
| 2 | ALICE | LS2 | 29 | 18.2 |
| 5 | CMS | LS1 | 29 | 21.7 |
| 8 | LHCb, VELO | LS2 | 5 | 3.5 |

**8.4   The passive forward absorbers**

The high luminosity regions of LHC at IP1 and IP5 are equipped with passive absorbers for charged (TAS) [5, 6] and neutral (TAN) [7] particles. The TAS is installed on either side of the interaction region at the transition of the experimental caverns to the LHC tunnel. TAS's and TAN's primary function is to protect the superconducting quadrupoles in the straight section from the collision debris coming from the interaction region. The TAS primarily shields the inner triplet quadrupoles Q1–Q3; TAN shields the D2 and Q4 quadrupoles. In parallel the TAS completes the forward shielding of the experiments and both the TAS and TAN participate in the reduction of background to the experiments.

For HL-LHC operation the following modifications are foreseen.

- New TAS and TAN absorbers on either side of IP1 and IP5, called TAXS and TAXN, respectively, should replace the existing ones. The protection must be extended to D1 magnets that in the HL-LHC will be superconducting (these are normally conducting in the present LHC). The new absorbers must have an aperture adapted to HL-LHC beam optics and operation, and should be designed to cope with the increased energy deposition.

- A new TAXN absorber is planned for IP8, designed to operate at the foreseen five times higher luminosity operation. The installation of a TAXS absorber around IP8 is not required.

- The new absorbers will be designed to operate at the ultimate luminosity conditions during the HL-LHC era as defined above.

8.4.1   The charged particle passive absorber – TAXS

The TAXS absorber is located approximately 19 m from the interaction point on either side of IP1 or IP5. Its core is a 0.5 m diameter and 1.8 m long copper cylinder traversed on its axis by a constant aperture beam pipe. The design of the new TAXS absorbers for the HL-LHC for IP1 and IP5 is based on that of the presently installed TAS absorbers, with the following modifications and improvements:

- the beam pipe aperture increases to 54 mm in diameter from the present 34 mm;

- the cooling power increases to dissipate approximately 780 W deposited in the TAXS during HL-LHC ultimate beam operation conditions, including a safety margin;

- the overall design of the TAXS remains compatible with the mechanical and envelope constraints from the surrounding shielding of the experiments.

Improvements in the alignment mechanism and vacuum exchange should be incorporated in the new design in view of the need for optimized maintenance operations and exposure to radiation. The change from the TAS to TAXS is to happen during LS3 after a few months (at least) of cool-down. The overall procedure must be optimized so as to minimize the exposure of personnel to radiation in compliance with the ALARA principle and the overall planning of the activities in the LHC tunnel and experimental caverns.

8.4.2   The neutral particle passive absorber – TAXN

The TAXN absorber is designed to absorb the flux of forward high-energy neutral particles produced at the interaction region of IP1 and IP5. There is a TAXN absorber installed in either side of IP1 and IP5, located between the separation/recombination dipole pair D1 and D2, containing the transition from the single common beam pipe to the two separate pipes for the incoming/outgoing beams.

The design of the new TAXN absorbers for IP1 and IP5 is based on that of the presently installed TAN, with the following modifications and improvements.

- The position of the TAXN is different by a few metres (towards the IP) with respect to the present position.

- The overall design and layout is to be adapted to the available space constraints while maintaining the correct shielding efficiency.
- The vacuum chamber layout is efficiently adapted to the new geometry for protection of the adjacent quadrupoles and optimization of the background conditions of the experiment. The TAXN vacuum pipes have a fixed aperture that, combined with a specially designed TCL collimator with movable jaws just downstream towards D2, provides the maximum protection efficiency at all beam optics scenarios for the HL-LHC.
- Active water cooling will be required to dissipate approximately 1.5 kW of power from the beam, expected during ultimate operation during the HL-LHC era.

Improvements in the mechanical design of the absorber should be incorporated in the design to allow optimized installation and maintenance activities. The locations for beam instrumentation for luminosity monitoring and experimental detectors will be maintained unless they impose important constraints in the TAXN design and maintenance.

In IP8 new TAXN absorbers should be installed on either side of the IR in available slots upstream of D2.